\documentclass[aps,prb,preprint,superscriptaddress]{revtex4-1}
\usepackage{graphicx}
\usepackage{dcolumn}
\usepackage{bm}
\usepackage{amssymb}
\usepackage{amsmath}
\usepackage{subfigure}
\usepackage{epstopdf}
\usepackage{float}

\begin{document}


\title{Continuous Nucleation Dynamics of Magnetic Skyrmions in T-shaped Helimagnetic Nanojunction}


\author{Ya-qing Zheng}
\affiliation{School of Physics, Nankai University, Tianjin 300071, China.}
\author{Yong Wang}
\email[]{yongwang@nankai.edu.cn}
\affiliation{School of Physics, Nankai University, Tianjin 300071, China.}


\begin{abstract}
Magnetic skyrmions are topologically-protected spin textures existing in helimagentic materials, which can be utilized as information carriers for non-volatile memories and logic circuits in spintronics. Searching simple and controllable way to create isolated magnetic skyrmions is desirable for further technology developments and industrial designs. Based on micromagnetic simulations, we show that the temporal dissipative structure can be developed in the T-shaped helimagnetic nanojunction when it is driven to the far-from-equilibrium regime by a constant spin-polarized current. Then the magnetic skyrmions can be continuously nucleated during the periodic magnetization dynamics of the nanojunction. We have systematically investigated the effects of current density, Dzyaloshinskii-Moriya interaction, external magnetic field, and thermal fluctuation on the nucleation dynamics of the magnetic skyrmions. Our results here suggest a novel and promising mechanism to continuously create magnetic skyrmions for the development of skyrmion-based spintronics devices.
\end{abstract}


\maketitle


\section{Introduction}
Magnetic skyrmions are nano-sized noncollinear magnetization configurations characterized by the topological skyrmion number,\cite{NatNano2013,NRM1,NRM2} which can exist in the noncentrosymmetric helimagnet with finite Dzyaloshinskii-Moriya interaction (DMI).\cite{Dzya,Moriya} Due to their topological stability and flexible controllablity by spin-polarized current, magnetic skyrmions show great advantages as information carriers to develop spintronics devices,\cite{Proc} such as memories,\cite{JPD} logic gates,\cite{SR2015} transistors,\cite{SR2015-2} \emph{etc}. For the successful operations of these skyrmion-based devices, the first step is to create isolated magnetic skyrmions in a simple and controllable way. Physically, the generation of an isolated mgnetic skyrmion implies the topological transition of magnetization configurations, which is accompanied by overcoming a potential barrier. Up to date, several methods to create isolated magnetic skyrmions from the uniform ferromagnetic(FM) states have been proposed and realized, including pulsed magnetic field,\cite{PRB2017} spin wave,\cite{APL2015} local heating,\cite{NC2014}, laser pulse,\cite{PRL2013} electric field,\cite{NatNano2017} and spin-polarized current.\cite{Science2013,NatNano2013-2,NatNano2013-3,PRB2012,APL2013,PRB2016,SR2016,NL2017} To avoid a brute-force destruction of the FM state, the effect of geometry confinement on the nucleation dynamics of magnetic skyrmions can be utilized.\cite{NatNano2013-3} Furthermore, it has been shown that the isolated magnetic skyrmion can be converted from either a domain-wall(DW) pair\cite{NC2014-2} or a chiral stripe domain\cite{Science2015} by spin torque in a junction geometry. 

\begin{figure*}[!ht]
  \centering
  \includegraphics[scale=0.5]{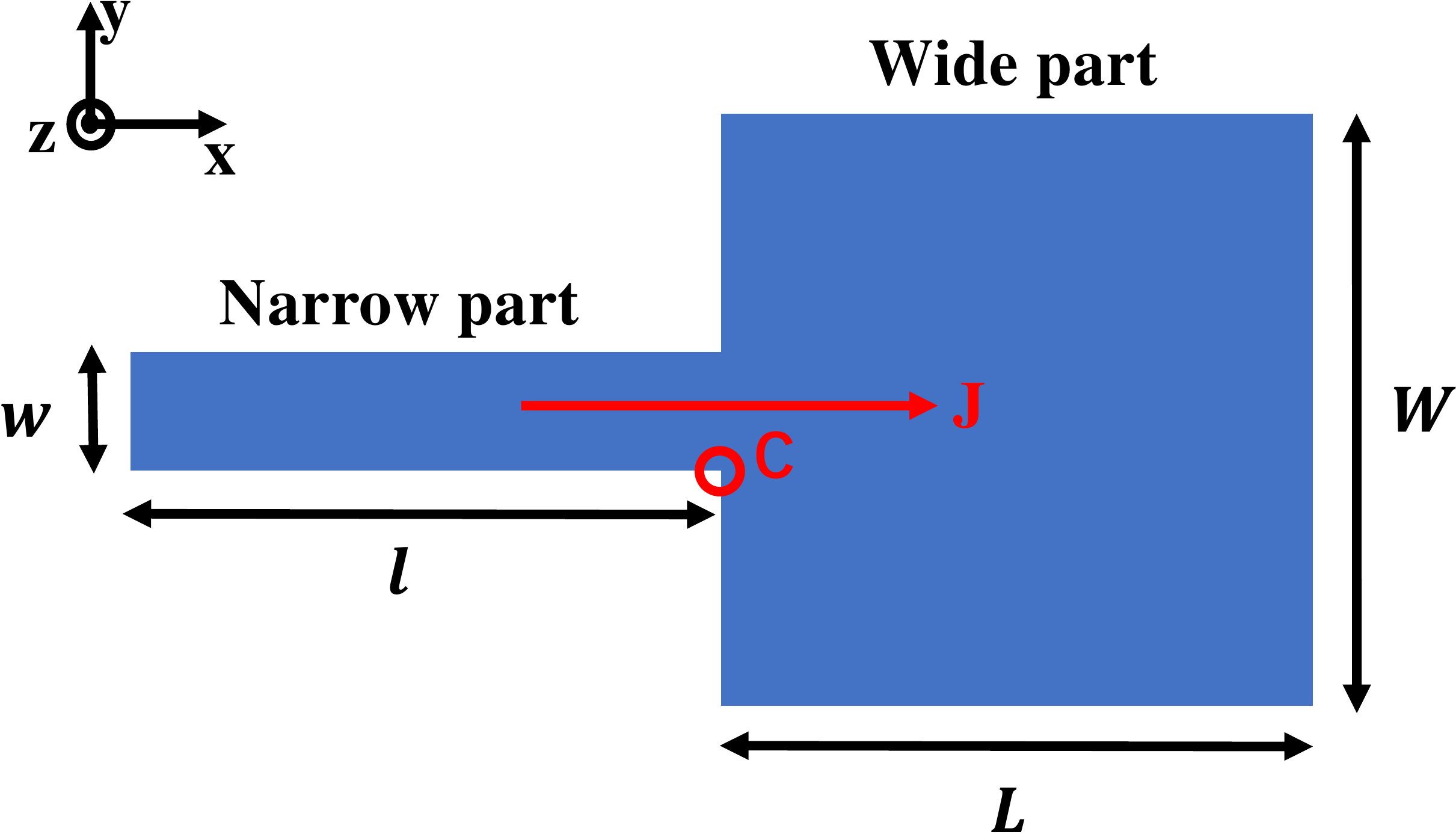}
  \caption{Schematic diagram of the T-shaped helimagnetic nanojunction from top view. The junction consists of a narrow part and a wide part, and an adjunctive layer (not shown explicitly) carrying a spin-polarized current $J$ along the $x$-direction. $L(l)$ : the length of the wide (narrow) part; $W(w)$ : the width of the wide (narrow) part. $C$: the point at the lower-right corner of the narrow part.} 
  \label{Fig1}
\end{figure*}

In this paper, we report a novel mechanism to continuously generate magnetic skyrmions in the T-shaped helimagnetic nanojunction based on micromagnetic simulations. When the nanojunction is driven far from the equilibrium by applying a spin-polarized current, the periodic magnetization dynamics can arise for certain simulation parameters, which is known as ``temporal dissipative structure".\cite{Dissipative2015} Since one magnetic skyrmion can be formed during each cycle of the dynamics, the magnetic skyrmions will be continuously generated by the temporal dissipative structure realized here. Unlike the previous mechanisms to generate magnetic skyrmions in the junction structure,\cite{NC2014-2,Science2015} where a DW pair or a stripe domain should be prepared \emph{in priori}, the ingredients in our mechanism here only include the special geometry shape of the nanojunction and the suitable spin-polarized current.             

\section{Structure and Method}
The T-shaped helimagnetic nanojunction considered here is schematiclaly shown in Fig.~\ref{Fig1}, which consists of a narrow part and a wide part. The length and width of the wide(narrow) region of the junction are denoted by $L$($l$) and $W$($w$) respectively, and the layer thickness is denoted by $d$. The average energy density of the helimagnet includes the Heisenberg exchange interaction (HEI), the DMI, the perpendicular magnetic anisotropy, and the Zeeman energy, which is explicitly defined as\cite{Proc}
\begin{eqnarray}
\mathcal{E}[\mathbf{M}]=A(\nabla\mathbf{m})^{2}+E_{DM}-K_{u}(\mathbf{e}_{z}\cdot\mathbf{m})^{2}-M_{s}\mathbf{B}\cdot\mathbf{m}.\label{Ham}
\end{eqnarray}
Here, $M_{s}$ is the saturation magnetization and $\mathbf{m}(\mathbf{r})=\mathbf{M}(\mathbf{r})/M_{s}$ is the normalized magnetization distribution, $A$ is the exchange stiffness, $K_{u}$ is the perpendicular magnetic anisotropy coefficient, $\mathbf{e}_{z}$ is the anisotropic axis, and $\mathbf{B}$ is the external magnetic field. Depending on the microscopic origin, the form of $E_{DM}$ can be either bulk DMI with $E_{DM}=D\mathbf{m}\cdot[\nabla\times\mathbf{m}]$ or interfacial DMI with $E_{DM}=D[m_{z}\nabla\cdot\mathbf{m}-(\mathbf{m}\cdot\nabla)m_{z}]$,\cite{Proc} where $D$ is the strength of DMI. Due to the competition between DMI and HEI, the helimagnetic materials can host Bloch-type skyrmion for bulk DMI and N\'eel-type skyrmion for interfacial DMI.\cite{Proc}

\begin{table*}[ht!]
\renewcommand\arraystretch{1.5}
\begin{center}
\caption{Geometry size and material parameters for the T-shaped helimagnetic nanojunction in the micromagnetic simulations. $L(l)$: length of the wide (narrow) part of the junction; $W(w)$: width of the wide (narrow) part of the junction; $d$: thickness of the junction; $M_{s}$: saturation magnetization; $A$ : exchange stiffness; $K_{u}$ : perpendicular magnetic anisotropy coefficient; $\alpha$ : Gibert damping cofficient.}\label{Table1}
\begin{tabular}{ccccccccc}
\hline\hline
$L$(nm) & $l$(nm) & $W$(nm) & $w$(nm) &$d$(nm) & $M_{s}$(kA/m) & $A$(pJ/m) & $K_{u}$(MJ/$\text{m}^{3}$) & \quad$\alpha$\quad \\
\hline
100 & 100 & 100 & 20 & 0.4 & 580 & 15 & 0.8 & \quad 0.3\quad \\
\hline\hline
\end{tabular}
\end{center}
\end{table*}

In our strategy, an adjunctive layer that can carry a spin-polarized current is placed on the helimagnetic layer. When the spin-polarized current is injected along the $x$-direction, the spin torque effect will be exerted on the helimagnet to excite the magnetization dynamics, which is described by the Landau-Lifshitz-Gilbert(LLG) equation\cite{NatNano2013-2}
\begin{eqnarray}
\frac{\emph{d}\mathbf{m}}{\emph{dt}}=-\gamma\mathbf{m}\times\mathbf{H}_{eff}+\alpha(\mathbf{m}\times\frac{\emph{d}\mathbf{m}}{\emph{dt}})
+\frac{u}{d}(\mathbf{m}\times\mathbf{m}_{p}\times\mathbf{m})
-\xi\frac{u}{d}\mathbf{m}\times\mathbf{m}_{p}.\label{LLG}
\end{eqnarray}
Here, $\gamma$ is the gyromagnetic ratio of electron, $\mathbf{H}_{eff}=-\frac{\partial\mathcal{E}[\mathbf{M}]}{\partial\mathbf{M}}$ is the effective magnetic field determined by the energy density functional $\mathcal{E}[\mathbf{M}]$, $\alpha$ is the Gilbert damping coefficient. The last two terms in Eq.~(\ref{LLG}) describe the damping-like and field-like torque.\cite{NatNano2013-2} The spin-polarized current is characterized by the current density $J$, the spin polarization degree $P$, and the spin polarization direction $\mathbf{m}_{p}$. The spin torque coefficient $u$ is given by\cite{PRB2013} $u=|\frac{\gamma\hbar JP}{2eM_{s}}|$, where $\hbar$ denotes the reduced Planck constant and $e$ is the elementary charge. The dimensionless parameter $\xi$ is adjusted to set the ratio between the field-like torque and the damping-like torque.\cite{NatNano2013-2} 

The restriction on spin polarization direction $\mathbf{m}_{p}$ to develop the temporal dissipative structure depends on the type of DMI. Here, we consider the continuous generation of N\'eel-type magnetic skyrmions, which can be realized by the spin-polarized current with $\mathbf{m}_{p}=-\mathbf{e}_{x}$. For Bloch-type magnetic skyrmions, a spin-polarized current with $\mathbf{m}_{p}=\mathbf{e}_{y}$ is required, and the similar phenomena can also be observed (see supporting information for details). The micromagnetic simulations for the magnetization dynamics are performed with the finite difference OOMMF code\cite{OOMMF}. The material parameters for Co/Pt films have been exploited,\cite{NatNano2013-2} which is listed in Table \ref{Table1} together with the geometry size of the nanojunction. To take into account the effect of field-like torque, which acts as an effective magnetic field along the $\mathbf{m}_{p}$ direction, we set the ratio $\xi=0.75$ during all the simulations. The imperfection of spin polarization is also considered by setting $P=0.4$. More details about the calculations can be found in the supporting information. 

\section{Results and discussion}
\subsection{Nucleation Dynamics of Single Magnetic Skyrmion}
\begin{figure*}[ht!]
\centering
  \includegraphics[scale=0.5]{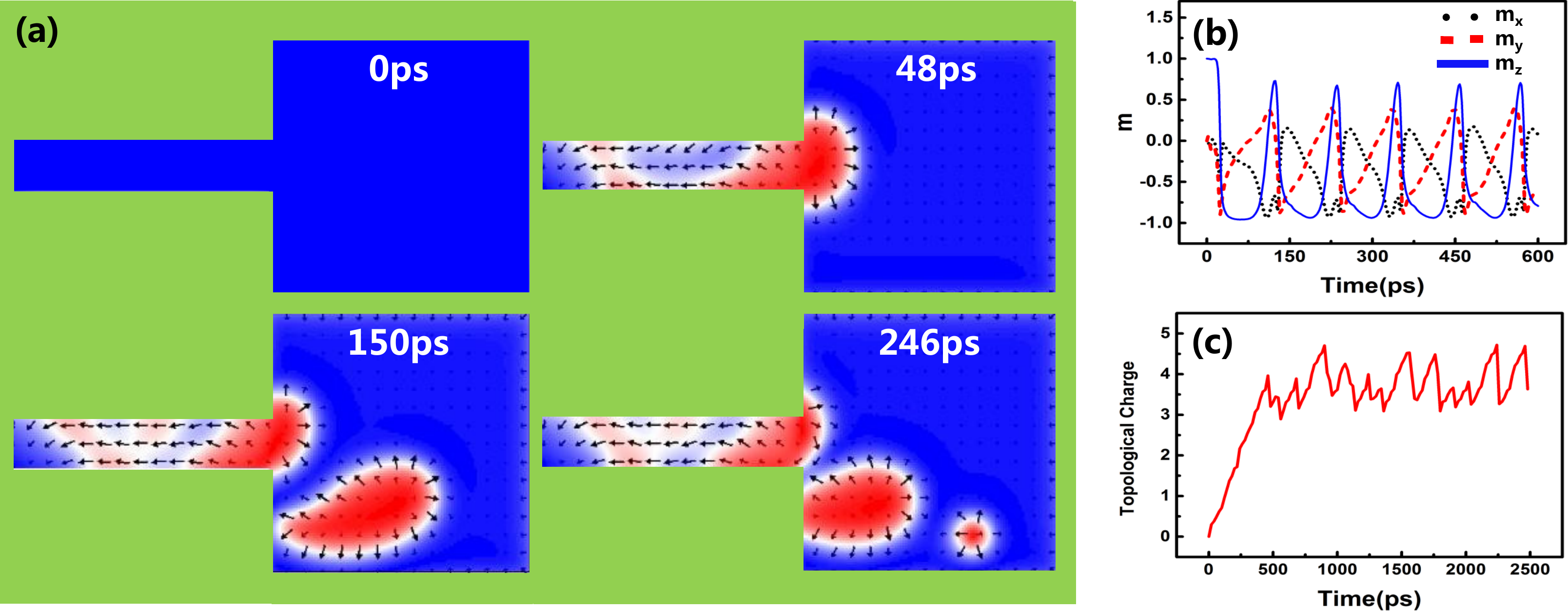}
  \caption{(a) Snapshots of the nucleation dynamics of a magnetic skyrmion driven by spin-polarized current in the T-shaped helimagnetic nanojunction at different time. (b) The oscillatory dynamics of the normalized magnetic moment $\mathbf{m}$ at corner point $C$. (c) The time-evolution of the topological charge in the whole region of the junction.}
  \label{Fig2}
\end{figure*}
We first demonstrate the nucleation dynamics of one magnetic skyrmion in the T-shaped nanojunction in Fig.~\ref{Fig2}(a). The DMI strength is set as $D=3.8$~mJ/m$^{2}$ and the current density at the narrow region is set as $J=2.0\times10^{12}$~A/m$^{2}$. The external magnetic field is not applied at the moment. Before the spin-polarized current is applied, the magnetization direction is uniformly aligned along the $\mathbf{e}_{z}$-direction due to the perpendicular magnetic anisotropy energy($t=0$~ps). When the current is applied, the magnetization at the narrow region of the junction will first be driven out of equilibrium, most of which will align along the $-\mathbf{e}_{x}$ direction due to the spin torque effect. This effect is however significantly weak in the wide region, where the current density is only $J/5$ due to the larger cross-sectional area. Nevertheless, the magnetization near the lower-right corner of the narrow region will rotate from $\mathbf{e}_{z}$ to $-\mathbf{e}_{z}$ direction, which can be regarded as a seed for the magnetic skyrmion. Because of the spin torque effect, the size of the ``skyrmion seed" will gradually expand in the wide region of the junction($t=48$~ps). Later on, the ``skyrmion seed" will be detached from the narrow region when the magnetization near its lower-right corner rotates back to $\mathbf{e}_{z}$($t=150$~ps). The isolated skyrmion seed then will continue to move in the wide region of the junction under the spin torque effect, during which its shape will gradually evolve and shrink down to a N\'eel-type magnetic skyrmion($t=246$~ps). The motion of the formed magnetic skyrmion will obey the Thiele equation\cite{Proc} until it is annihilated at the boundary of the junction. The same dynamical process will happen repeatly, which then results in the continuous nucleation of the magnetic skyrmions in the T-shaped nanojunction. 

For a better understanding of the dynamical process described above, we investigate the magnetization dynamics at the corner point $C$ of the narrow region. As shown in Fig.~\ref{Fig2} (b), the magnetic moment at point $C$ will oscillate with a period $T_{0}=112$~ps, which is accompanied by the repeating nucleation dynamics of magnetic skyrmions. In fact, it is the collective dynamics of all the oscillatory magnetic moments in the nanojunction that gives rise to the temporal dissipative structure. Furthermore, we numerically calculate the time-evolution of the topological skyrmion number $Q=-\frac{1}{4\pi}\int\textbf{m}\cdot(\frac{\partial\textbf{m}}{\partial x}\times\frac{\partial\textbf{m}}{\partial y})d^{2}\mathbf{r}$,\cite{NatNano2013,Proc} which is integrated over the whole area of the junction. For the simulated magnetization dynamics here, the skyrmion number will first increase from $0$ to $4$, and then oscillate between the values $3$ and $4.7$ ( Fig.~\ref{Fig2}(c)), which corresponds to the creation and annihilation of magnetic skyrmions. Due to the finite size of the junction, the maximal number of coexisting magnetic skyrmions in the temporal dissipative structure here is $4$, and the extra $0.7$ comes from the partially-formed skyrmion seed.       

Besides, we found that the evolution dynamics of the skyrmion seeds can be complicated by the skyrmion-skyrmion interaction or skyrmion-edge interaction in the wide region of the junction. For example, the evolution from a skyrmion seed to a magnetic skyrmion requires enough space. When the skyrmion seed gets close to the bottom boundary of the junction, it may quickly evolve to a magnetic skyrmion due to the skyrmion-edge interaction. On the other hand, the distance between the magnetic skyrmions and their sizes can vary with time due to skyrmion-skyrmion interaction. If two magnetic skyrmions get close, both of them will shrink down due to the repulsive interaction; then their distance will get longer and their sizes will become large again. Besides, the motion of magnetic skyrmions will get slower near the edge, such that the following magnetic skyrmion will catch up with the former one and compress it through the skyrmion-skyrmion interaction.

\subsection{Dzyaloshinskii-Moriya interaction and Current Density}
Now we investigate the effect of DMI strength $D$ and current density $J$ on the nucleation dynamics of the magnetic skyrmions in the junction. The same geometry size and the material parameters have been set as above. By sweeping the DMI strength $D\in[3.4,4.9]$~mJ/m$^{2}$ and the current density $J\in[17,26]$~A/m$^{2}$, we find that the magnetization dynamics in the junction can be classified as the following cases (see supporting information for more details):

   (I) no skyrmion seed can be formed in the the junction for small current density;
           
   (II) all the skyrmion seeds will move along the left boundary and fail to form magnetic skyrmions;
   
   (III) some but not all the skyrmion seeds can evolve into the magnetic skyrmions; 
   
   (IV) all the skyrmion seeds will evolve into the magnetic skyrmions continuously, which is the case we are interested in; 
   
   (V) all skyrmion seeds will move and stick to the bottom boundary of the wide region, and no magnetic skyrmion can be formed.  
   
\begin{figure*}[ht!]
\centering
  \includegraphics[scale=0.5]{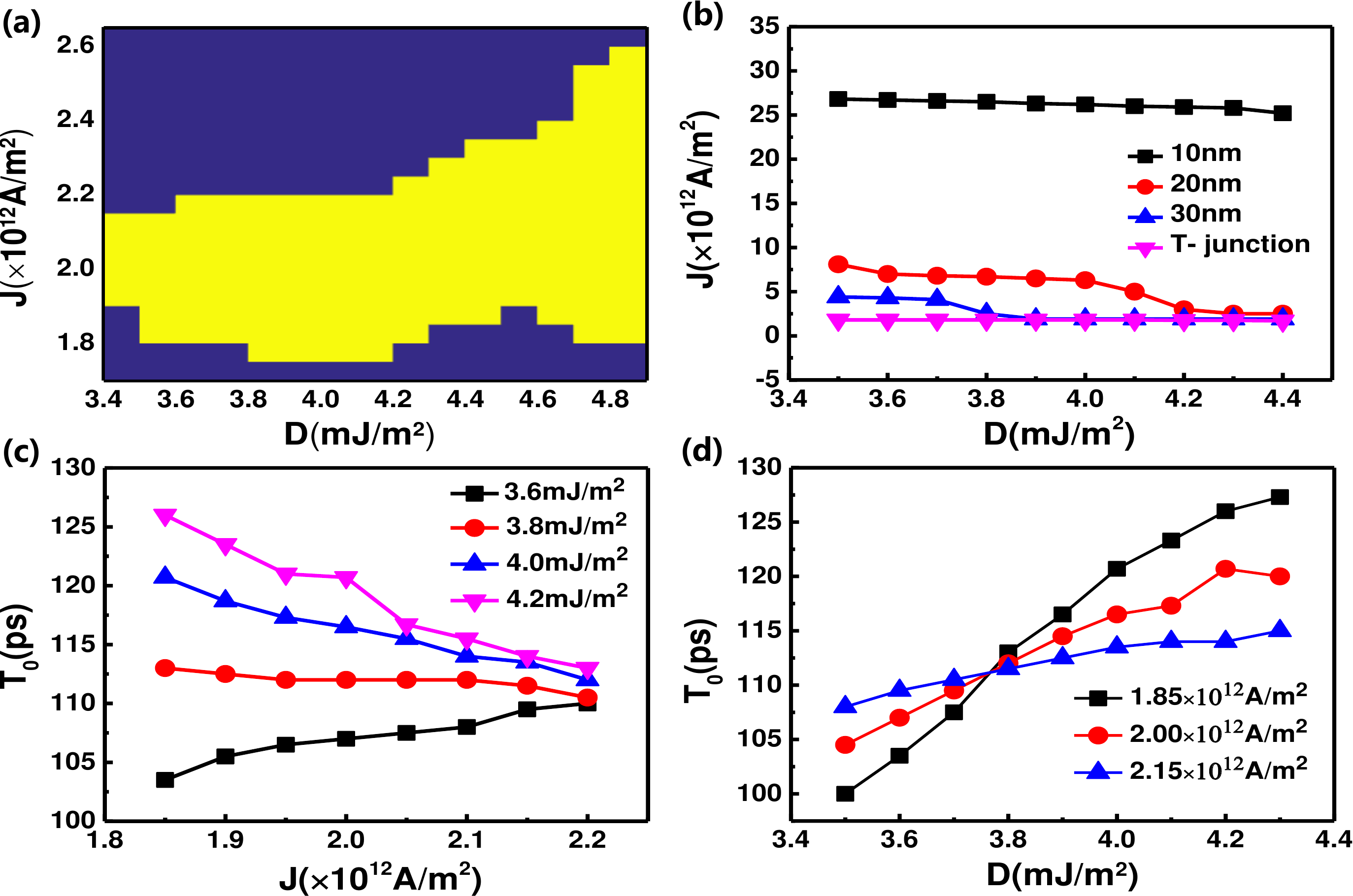}
  \caption{ (a) The $D$-$J$ phase diagram. The yellow region gives the parameter space for which the continuous nucleation dynamics of magnetic skyrmions can happen. (b) The critical current densities to generate magnetic skyrmions in the current-perpendicular-to-plane (CPP) configuraton and the T-shaped nanojunction (T-junction). Three different diameters $10,20,30$~nm have been set for the contact region in CPP configuration. (c) The dependence of nucleation period $T_{0}$ on the injected current density $J$ for different DMI strength $D$. (d) The dependence of nucleation period $T_{0}$ on the DMI strength $D$ for different injected current density $J$.}
  \label{Fig3}
\end{figure*}

The parameter space for case (IV) is displayed in the $D$-$J$ phase diagram in Fig.~\ref{Fig3}(a). For each value of DMI strength $D$, there exist a lower critical current density $J_{c1}$ which triggers the temporal dissipative structure to continuously create magnetic skyrmions, and an upper critical current density $J_{c2}$ beyond which the dynamics will vanish. Therefore, the continuous formation of magnetic skyrmions from the skyrmion seeds can only happen with a moderate current density. From the $D$-$J$ phase diagram, we see that $J_{c1}$ slightly depends on the DMI $D$, since the energy barrier to form a magnetic skyrmion mainly depends on the Heisenberg exchange interaction. In contrast, $J_{c2}$ becomes larger along with the increasement of $D$, since the formed magnetic skyrmion will be more stable with stronger DMI.  
In the conventional current-perpendicular-to-plane(CPP) configuration,\cite{NatNano2013-3} where the spin-polarized current is vertically injected to the helimagnetic layer to create magnetic skyrmions, the spin torques take the same forms as Eq.~(\ref{LLG}). However, we expect that higher current density is required to overcome the energy barrier to create an isolated magnetic skyrmion, since the geometry confinement is not utilized in this configuration. In Fig.~\ref{Fig3}(b), we show the critical current density $J_{c}$ to create a magnetic skyrmion in the wide region with the CPP method. We see that $J_{c}$ will gradually decrease when the DMI strength $D$ is increased for fixed contact area. While for given $D$, the critical density $J_{c}$ will significantly decrease when the diameter of the contact region is increased from $10$~nm to $30$~nm. For comparison, we also show the critical current density $J_{c1}$ to continuously create magnetic skrymions in the T-shaped junction, which is lower than the critical current densities in CPP configuration here. Therefore, the mechanism proposed here has the advantage to create magnetic skyrmions with lower current density.  

Based on the $D$-$J$ phase diagram, we further investigate the effect of current density $J$ and the DMI strength $D$ on the period $T_{0}$ of the continuous nucleation dynamics in case (IV), which is retrieved from the oscillation of magnetic moment at the corner point $C$ as in Fig.~\ref{Fig2}(b). As shown in Fig.~\ref{Fig3}(c), when the current density $J$ is increased from $18.5$~A/m$^{2}$ to $22$~A/m$^{2}$, the range of nucleation period $T_{0}$ will become narrower for $D=3.6,3.8,4.0,4.2$~mJ/m$^{2}$. Thus, the DMI will become less importance on the nucleation period  when the current density $J$ approaches the upper critical value $J_{c2}$. This feature becomes more transparent in the $T_{0}$-$D$ relations for different current density $J$, as shown in Fig.~\ref{Fig3}(d). Here, the slope of $T_{0}$-$D$ curve will decrease when the current density $J$ increases. Another feature in Fig.~\ref{Fig3}(c) is that the nucleation period $T_{0}$ is almostly independent on the current density $J$ when $D\approx 3.8$~mJ/m$^{2}$, which results in a cross region near $D\approx 3.8$~mJ/m$^{2},T\approx 110$~ps for the $T_{0}$-$D$ curves in Fig.~\ref{Fig3}(d). Therefore, when the current density is increased, the nucleation period $T_{0}$ will decrease if $D>3.8$~mJ/m$^{2}$ but increase if $D<3.8$~mJ/m$^{2}$, as given in Fig.~\ref{Fig3}(c). The results here imply that the nucleation dynamics of the magnetic skyrmions in the T-shaped nanojunction can be tuned by the injected current density.    

\subsection{External Magnetic Field}
\begin{figure*}[ht!]
\centering
  \includegraphics[scale=0.5]{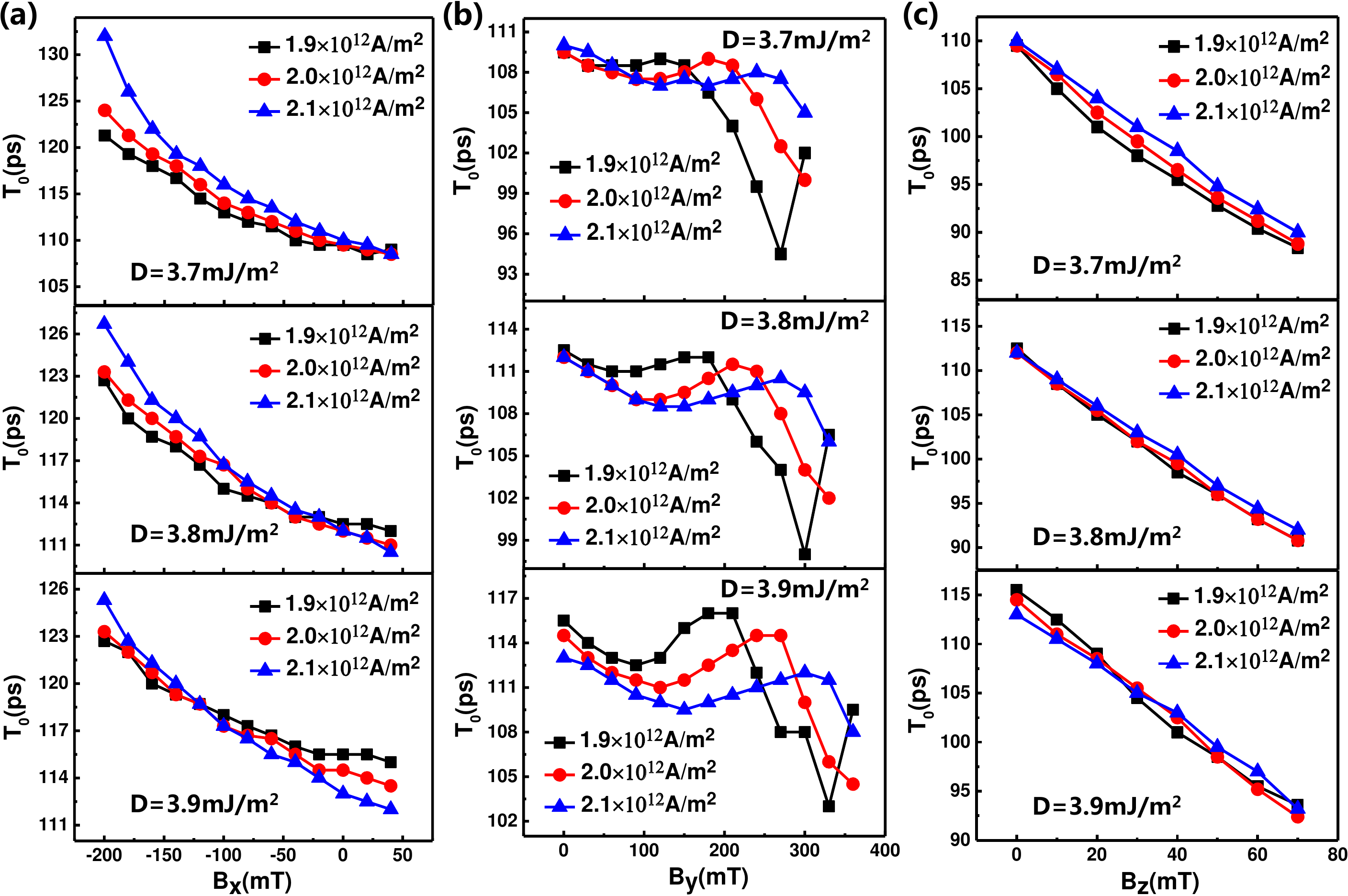}
  \caption{The response of nucleation period $T_{0}$ to the external magnetic field $\mathbf{B}$ for different DMI strength $D$ and injected current density $J$. The direction of $\mathbf{B}$ is set as : (a) $x$; (b) $y$; (c) $z$ respectively.}
  \label{Fig4}
\end{figure*}

We further investigate the response of the continuous nucleation dynamics of magnetic skyrmions in the presence of external magnetic field. When the amplitude of magnetic field is very large and the Zeeman term is dominant over the other contributions in Eq.~(\ref{Ham}), we expect that the nucleation dynamics will be suppressed or even eliminated. For several typical parameter sets $\{D,J\}$ with $D=3.7,3.8,3.9$~mJ/m$^{2}$ and current density $J=1.9\times10^{12},2.0\times10^{12},2.1\times10^{12}$~A/m$^{2}$ in the $D$-$J$ phase diagram, we sweep the magnetic field in the $\mathbf{e}_{x}$, $\mathbf{e}_{y}$, and $\mathbf{e}_{z}$ direction respectivley, and retrieve the period $T_{0}$ of the nucleation dynamics. We find that the continuous nucleation dynamics can robustly exist in a rather wide range of external magnetic field, and the dependence of nucleation period $T_{0}$ on the magnetic field $\mathbf{B}$ is shown in Fig.~\ref{Fig4}.
Firstly, when the external magnetic field $\mathbf{B}=(B_{x},0,0)$ is applied, the continuous nucleation dynamics will exist if $B_{x}\in[-200,50]$~mT for all the parameter sets. The nucleation period $T_{0}$ can increase exponentially by tens of picoseconds when the magnetic field reaches $B_{x}=-200$~mT. Besides, the amplitude of increasement is larger for higher current density $J$, namely, the nucleation dynamics driven by higher current density is more sensitive to the magnetic field along $\mathbf{e}_{x}$-direction. Secondly, when the external magnetic field is applied as $\mathbf{B}=(0,B_{y},0)$, its range to keep the continuous nucleation dynamics for all the parameter sets is $B_{y}\in[0,300]$~mT. The nucleation period $T_{0}$ shows a pattern of oscillatory decay when $B_{y}$ is increased, and the variation of $T_{0}$ is less than $20$~ps. In contrast to the magnetic field along $\mathbf{e}_{x}$ direction, the nucleation dynamics is more sensitive to the magnetic field along $\mathbf{e}_{y}$ direction when the current density is lower. Lastly, the condition to keep the continuous nucleation dynamics for the magnetic field $\mathbf{B}=(0,0,B_{z})$ is $B_{z}\in[0,70]$~mT for all the parameter sets. Here, the nucleation period $T_{0}$ will decay linearly when the $B_{z}$ is increased, and the variation of $T_{0}$ is less than $30$~ns. Besides, the slope of $T_{0}-B_{z}$ curve shows less dependence on the current density.
Therefore, the continuous nucleation dynamics will response in different way when the external magnetic field is applied in different direction, which can also be used to tune the nucleation dynamics of magnetic skyrmions.

\begin{figure*}[ht!]
\centering
  \includegraphics[scale=0.5]{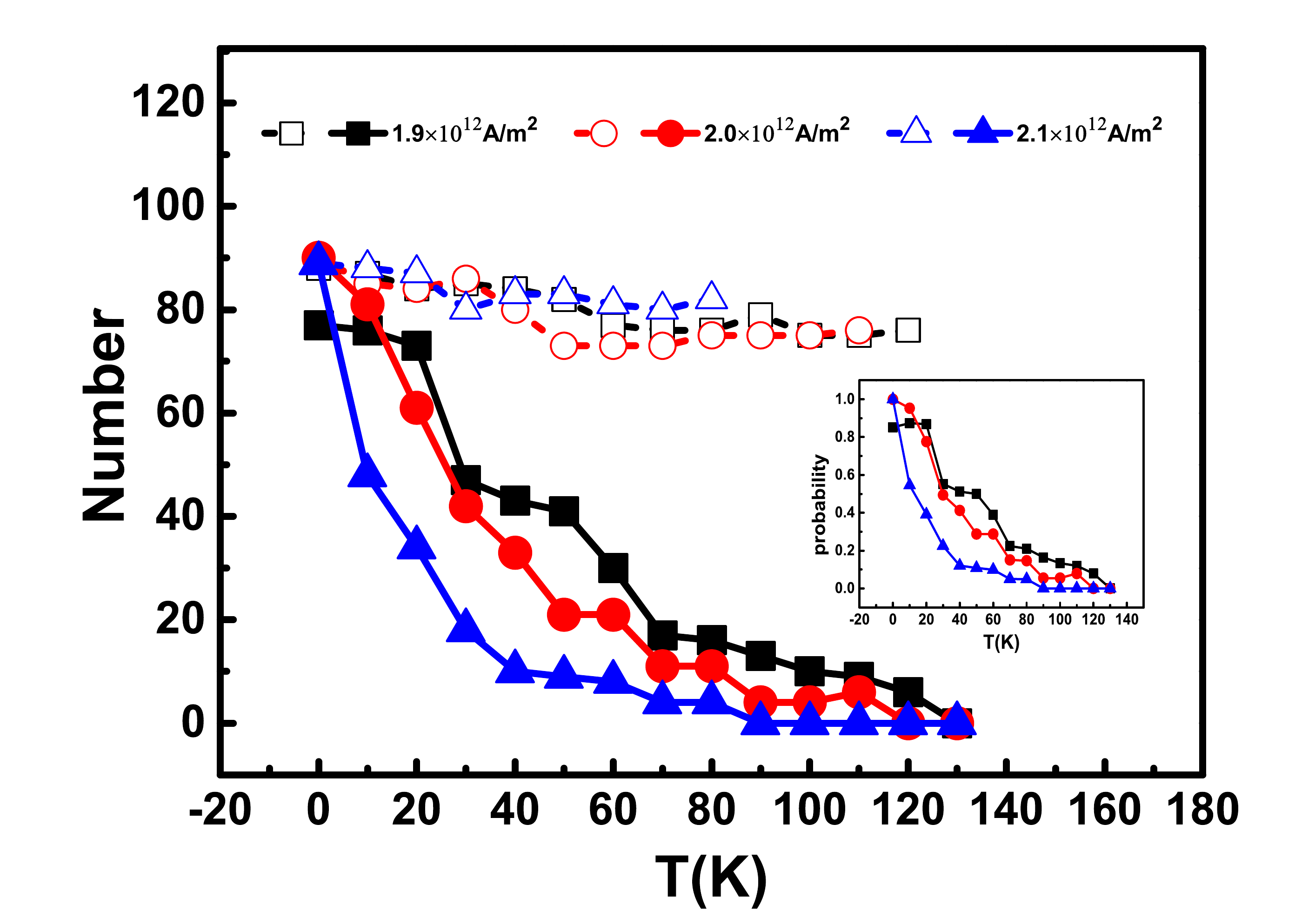}
  \caption{The dependence of number of skyrmion seed $N_{seed]}$ and magnetic skyrmion $N_{skyr}$ on the temperature $T$ for three different injected current density. Here, the DMI strength $D=3.8$~mJ/m$^{2}$. Inset: the probability $p=N_{skyr}/N_{seed}$ to successfully generate magnetic skyrmions at different temperature.}
  \label{Fig5}
\end{figure*}
\subsection{Thermal Stability}
Finally, we investigate the thermal stability of the continuous nucleation dynamics of magnetic skyrmions at finite temperatures. When the thermal fluctuation is present, some of the skyrmion seeds will not evolve into magnetic skyrmions, but will be destroyed and then disappear at the bottom boundary of the wide region in the junction (see supporting information for more details). Fig.~\ref{Fig5} shows the numbers of skyrmion seeds $N_{seed}$ and successfully formed magnetic skyrmions $N_{skr}$ during the nucleation dynamics for $10$~ns at different temperature, where the DMI strength is fixed as $D=3.8$~mJ/m$^{2}$ and three different current densities $J=19,20,21$~A/m$^{2}$ have been chosen. We see that $N_{seed}$ doesn't obviously depend on the temperature, but $N_{skr}$ quickly decays when the temperature is higher. The dependence of ratio $p=N_{skr}/N_{seed}$ on the temperature is further shown as inset in Fig.~\ref{Fig5}. At low temperatures, most of the skyrmion seeds will evolve into magnetic skyrmions and $p$ is close to $1$. For higher temperatures, the probability to form magntic skyrmions will decrease and finally become zero at a critical temperature. For fixed temperature, the skyrmion seeds are more easily destroyed and $p$ gets smaller if the current density is larger, which implies a smaller critical temperature above which no magnetic skyrmion can be generated. Besides, the critical temperatures are far below the Curie temperature, which is about $220$~K as given in the supporting information.  

\section{Conclusion}

In conclusion, we have revealed that the magnetic skyrmions can be continuously created by spin torque effect in the T-shaped helimagnetic nanojunction. Here, the magnetization dynamics of the nanojunction is driven to the far-from-equilibrium regime by spin-polarized current, and the temporal dissipative structures can be developed. In certain parameter space, one magnetic skyrmion can be generated during per cycle of the periodic magnetization dynamics. The nucleation period of the magnetic skyrmions can be further tuned by current density and external magnetic field, and the nucleation dynamics can be destroyed by thermal fluctuations. The discovery here offers a simple and controllable mechanism to continuously create magnetic skyrmions, which can have potential applications to develop skyrmion-based spintronics devices at nanoscale. Furthermore, it is also valuable to understand how the topological defects can be continuously generated in the far-from-equilibrium dynamics in general. 

\section*{Acknowledgements}
This work is supported by NSFC Project No. 61674083 and No. 11604162.

\bibliography{skybib}

\end{document}